\RequirePackage[hyphens]{url}
\documentclass[sigconf, authorversion=true, nonacm=true]{acmart}
\usepackage{algorithm}
\newcommand{\pluseq}{\mathrel{+}=}
\usepackage{array}
\usepackage{booktabs}
\usepackage[tight,footnotesize]{subfigure}

\usepackage{listings}
\usepackage{color}

\definecolor{mygreen}{rgb}{0,0.6,0}
\definecolor{mygray}{rgb}{0.5,0.5,0.5}
\definecolor{mymauve}{rgb}{0.58,0,0.82}

\lstdefinelanguage{JavaScript}{
  morekeywords=[1]{break, continue, delete, else, for, function, if, in,
    new, return, this, typeof, var, void, while, with},
  morekeywords=[2]{false, null, true, boolean, number, undefined,
    Array, Boolean, Date, Math, Number, String, Object},
  morekeywords=[3]{eval, parseInt, parseFloat, escape, unescape},
  sensitive,
  morecomment=[s]{/*}{*/},
  morecomment=[l]//,
  morecomment=[s]{/**}{*/}, 
  morestring=[b]',
  morestring=[b]"
}[keywords, comments, strings]

\newif\ifdraft
\draftfalse 

\begin{document}

\title{Eradicating the Unseen: Detecting, Exploiting, and Remediating a Path Traversal Vulnerability across GitHub}

\author{Jafar Akhoundali}
\affiliation{%
  \institution{LIACS, Leiden University}
  \city{Leiden} 
  \country{The Netherlands} 
}
\email{j.akhoundali@liacs.leidenuniv.nl}

\author{Hamidreza Hamidi}
\affiliation{%
  \institution{Technical and Vocational University}
  \city{Mashhad}  
  \country{Iran}
}
\email{0xxparrot@gmail.com}

\author{Kristian Rietveld}
\affiliation{%
  \institution{LIACS, Leiden University}
  \city{Leiden} 
  \country{The Netherlands} 
}
\email{k.f.d.rietveld@liacs.leidenuniv.nl}

\author{Olga Gadyatskaya}
\affiliation{%
  \institution{LIACS, Leiden University}
  \city{Leiden} 
  \country{The Netherlands} 
}
\email{o.gadyatskaya@liacs.leidenuniv.nl}





\begin{abstract}

Vulnerabilities in open-source software can cause cascading effects in the modern digital ecosystem. It is especially worrying if these vulnerabilities repeat across many projects, as once the adversaries find one of them, they can scale up the attack very easily. Unfortunately, since developers frequently reuse code from their own or external code resources, some nearly identical vulnerabilities exist across many open-source projects.

We conducted a study to examine the prevalence of a particular vulnerable code pattern that enables path traversal attacks (CWE-22) across open-source GitHub projects. To handle this study at the GitHub scale, we developed an automated pipeline that scans GitHub for the targeted vulnerable pattern, confirms the vulnerability by first running a static analysis and then exploiting the vulnerability in the context of the studied project, assesses its impact by calculating the CVSS score, generates a patch using GPT-4, and reports the vulnerability to the maintainers. 

Using our pipeline, we identified 1,756 vulnerable open-source projects, some of which are very influential. For many of the affected projects, the vulnerability is critical (CVSS score higher than 9.0), as it can be exploited remotely without any privileges and critically impact the confidentiality and availability of the system. We have responsibly disclosed the vulnerability to the maintainers, and 14\% of the reported vulnerabilities have been remediated. 

We also investigated the root causes of the vulnerable code pattern and assessed the side effects of the large number of copies of this vulnerable pattern that seem to have poisoned several popular LLMs. Our study highlights the urgent need to help secure the open-source ecosystem by leveraging scalable automated vulnerability management solutions and raising awareness among developers.\footnote{To appear in ACM AsiaCCS 2025. This is the author-prepared version.}

\end{abstract}

\maketitle


\section{Introduction}

Open-source software is a cornerstone of software systems today. Academic and industry studies~\cite{li2024systematic,harutyunyan2020managing,blind2021impact,ossra2024} highlight the critical dependency of our digital ecosystems on open-source projects. However, this dependency also creates various security risks and, for example, enables software supply chain attacks~\cite{ladisa2023sok,ohm2020backstabber,ossra2024,zimmermann2019small}. 

One particular source of security concerns is that developers tend to reuse code, whether coming from their codebase or other projects and community sources~\cite{chen2024empirical,mokhberi2021makes}.
However, as programmers copy-paste the code, they can transfer vulnerabilities from that code into their own or create new vulnerabilities~\cite{kim2017vuddy,woo2022movery,woo2021v0finder,reid2022extent}. For example, it was shown that developers copy vulnerable code snippets from StackOverflow~\cite{fischer2017stack,verdi2020empirical,chen2024empirical}.
The proliferation of such vulnerable code patterns across many open-source projects multiplies the risks. First, the attackers need to know only one pattern to be able to successfully attack many projects and their downstream dependencies~\cite{yan2021estimating,reid2022extent}. Second, the widespread presence of a vulnerable pattern across many codebases may normalize this code to developers, and they will continue copy-pasting it further and will not be so confident in identifying it as vulnerable~\cite{mokhberi2021makes}. Additionally, today, in the age of large language models (LLMs) that are trained on open-source codebases~\cite{kocetkov2022stack,yang2024gotcha}, the presence of a repeated vulnerable code pattern across many projects enables further replication of the vulnerability by means of code generation~\cite{al2023ab,cotroneo2024vulnerabilities,trufflesec2024}.  

In this work, motivated by the knowledge of the widespread vulnerable code reuse, we conduct a study of one such vulnerable code pattern and its spread across GitHub, which is currently the most popular online platform to host open-source software projects. 

The focus of our research is a \emph{path traversal vulnerability} (CWE-22~\cite{cwe22}) that enables adversaries to access files or resources in the vulnerable system, which are outside of the restricted location designated for anonymous web users. This type of vulnerability is very dangerous, as it may threaten the confidentiality, integrity, and availability of the system. According to MITRE, this vulnerability was among the top 25 most dangerous software weaknesses for 2024~\cite{cwe2024top25}. It is also among the top 10 known exploited vulnerabilities (KEV) for 2024~\cite{cwe2024kev}, with an average CVSS score of 8.09 (high) for KEVs of this type. 

Within this research, we explore a particular path traversal vulnerable code pattern that we spotted in different Node.js projects, where developers seem to be unaware of its security implications. During our initial investigation, we identified the usage of this pattern in two prominent open-source projects, obtaining two CVEs for them, and questioned ourselves about how many more vulnerable open-source projects are out there. But this is a question that cannot be investigated manually due to the vast size of GitHub and the complexity of validating each single vulnerability instance and examining its security implications in the context of each project. At the same time, discovering a vulnerability creates an ethical obligation for security researchers to help the maintainers patch it. This is a challenging and time-consuming task in itself, as a project-specific patch needs to be developed and responsibly communicated to the maintainers. Thus, we decided to study the prevalence of this code pattern systematically, while also developing a means to remediate it as much as possible. The key objectives of our study are the following:

\begin{itemize}
\item[\textbf{O1:}] Examine the presence of this path traversal code pattern across open-source GitHub projects.
\item[\textbf{O2:}] Understand the security implications for vulnerable projects by checking the exploitability of the code and evaluating the CVSS score of the confirmed vulnerability instances.
\item[\textbf{O3:}] Raise awareness and help the maintainers to patch the code in a scalable yet responsible manner.
\item[\textbf{O4:}] Examine the root causes of the vulnerable code pattern.
\item[\textbf{O5:}] Assess the influence of the widespread vulnerable code pattern on the current popular LLMs.
\end{itemize}

\textbf{Our contributions are the following:}

(1) We developed a pipeline to detect the studied critical path-traversal vulnerability code pattern in open-source projects \textbf{at the GitHub scale} that automates the detection, exploitation, and remediation (including patch generation, CVSS score estimation, and reporting). Our pipeline is released open-source\footnote{The repo will be published after the paper is presented at AsiaCCS in \url{https://github.com/JafarAkhondali/DotDotDefender}}.

(2) Using our pipeline, we scanned GitHub and identified \textbf{1,756 exploitable instances of path traversal} that we started to responsibly report to the maintainers.  Some of these vulnerabilities were discovered in very popular projects with thousands of stars, sometimes hosted under the umbrella of prominent organizations. Eliminating this vulnerable pattern in open-source projects thus substantially contributes to the confidentiality and availability of software. Our prototype pipeline has already led to 63 projects being fixed, and more projects that have been informed and are in the process of fixing the issue. We report on the responses of the maintainers that we have received so far.

(3) We investigated the root causes of the vulnerable code snippet that first emerged around 2010 and evaluated its impact on LLMs. We traced how the code pattern migrated between different reputable community platforms and developer learning resources. We saw that developers voicing security concerns were not supported by the majority, reflecting prejudice~\cite{barzilay2014understanding} and indicating that more effort is needed to raise security awareness about such tricky vulnerabilities. 
The current LLM chatbots, however, are in a far worse situation: we show that they are contaminated with the studied vulnerable code pattern and generate it even when instructed to synthesize secure code. 

To the best of our knowledge, our study is the first one to both examine the prevalence of the critical path traversal vulnerable code pattern in open-source GitHub projects and provide an automated means to detect and remediate novel instances of this vulnerability at a large scale.

\begin{figure*}[!t]
  \centering
  \includegraphics[width=0.8\textwidth, keepaspectratio]{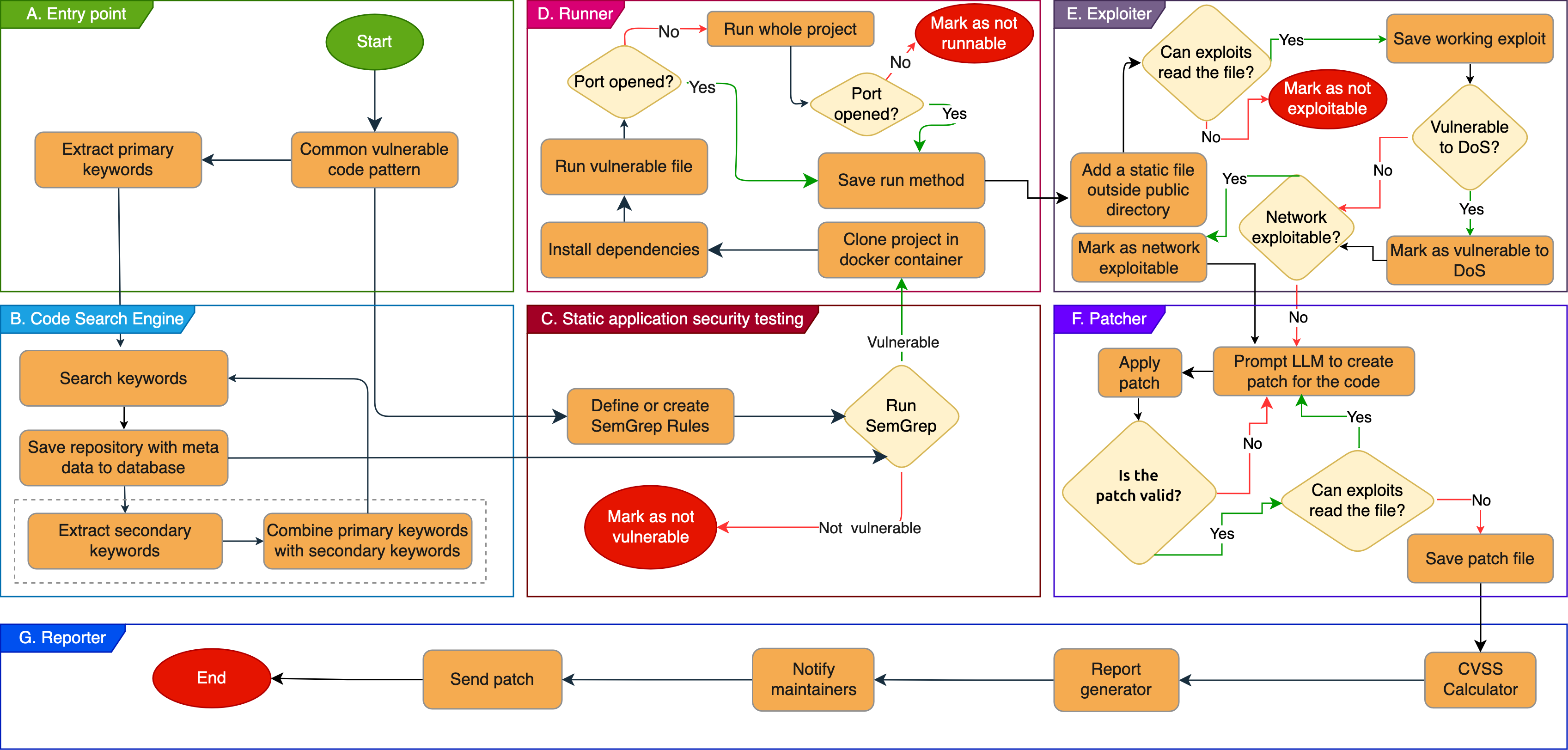}
  \caption{Overall flowchart of the proposed pipeline.}
   \label{fig:flow}
\end{figure*}

\section{Methodology}\label{sec:methodology}
In this section, we describe the methodology of our study. First, we discuss the path traversal vulnerable code pattern that is the focus of this study. To accomplish the objectives of our study, we need to detect, exploit, and remediate this vulnerability in a considerable number of open-source projects. This is only possible through automated means. Thus, in the remainder of this section, we describe an automated pipeline that achieves our objectives. Although in prior literature automated implementations of the independent steps of detection~\cite{go2023simple,jona22bh,shcherbakov2023silent,kang2023scaling,li2021detecting}, exploitation~\cite{kang2023scaling,cassel2023nodemedic,cassel2024nodemedic,wu2024autopt,fang2024llmexploit} and remediation~\cite{bui2024apr4vul,wang2020automated,fei2025patch,pearce2023examining,xia2023automated,wu2023effective,fu2022vulrepair,shen2020survey,pinconschi2021comparative} are described, our pipeline is -- to the best of our knowledge -- the first to automate the complete end-to-end process from detection to remediation (including responsible disclosure) of path traversal vulnerabilities in real-world Node.js projects at the GitHub scale.

\begin{figure*}
  \centering
  \includegraphics[width=0.78\textwidth, keepaspectratio]{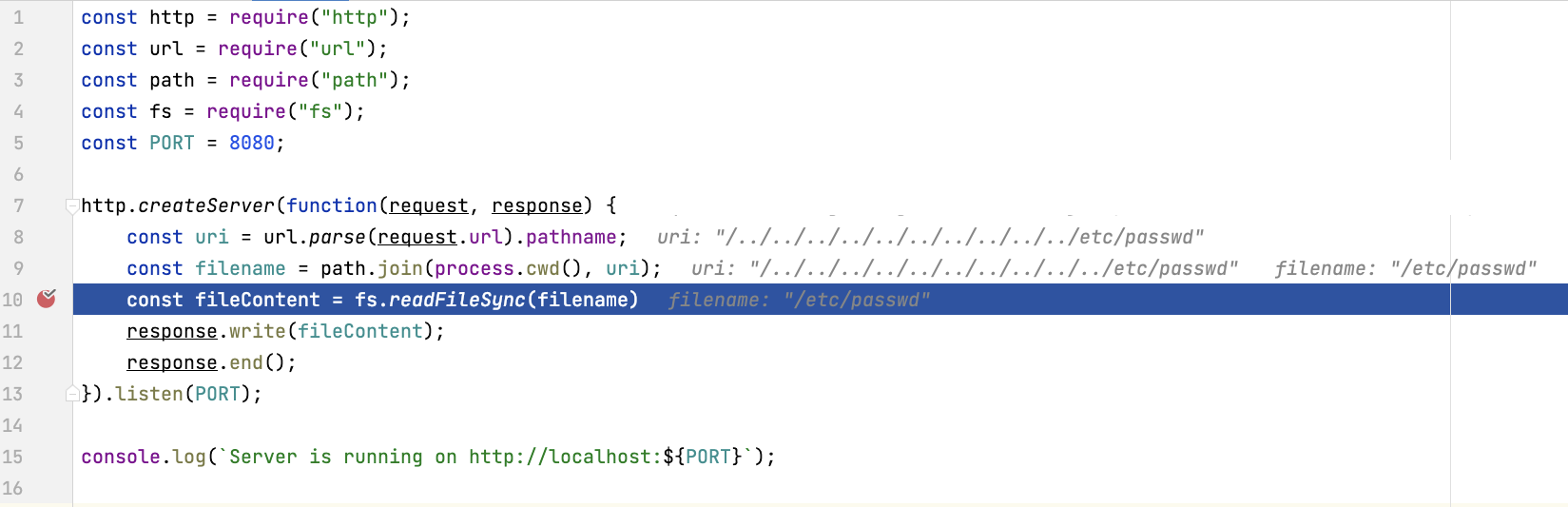}
  \caption{Simplified JavaScript (Node.js) code vulnerable to the path traversal attack.}
   \label{fig:vulnerablecode}
\end{figure*}

\subsection{Path Traversal Vulnerable Code Pattern}\label{sec:snippet}
The focus of this study is on a simple and popular code pattern that creates a static HTTP file server for Node.js web applications. A simplified version of this code pattern is shown in Fig.~\ref{fig:vulnerablecode}. Despite its simplicity and popularity, many developers appear unaware that this code pattern is vulnerable to the path traversal attack~\cite{cwe22}. The vulnerability manifests itself as follows. In line 8, the parser extracts the \texttt{pathname} provided by the user. Then, at line 9, the path of the current working directory (the public directory intended to be served to users) is joined using the \texttt{join} function. The \texttt{join} function first concatenates the two paths and then normalizes the final path. This allows attackers to provide a \texttt{pathname} that resolves to directories outside of the public directory intended to be served.

This pattern can be exploited in two ways. First, consider a current working directory of \texttt{/home/myapp/} and a user-supplied \texttt{pathname} of \texttt{../../etc/passwd}. This resolves to the final path \texttt{/etc/passwd}. Consequently, malicious users can access files outside of the directory intended to be served, thereby damaging the confidentiality of the system. This vulnerability also allows reading special UNIX-like files like \texttt{/proc/self/environ} that expose environment variables.
Second, this code snippet is vulnerable to denial of service attacks. When receiving a request to read a very large file (e.g., \texttt{/dev/urandom}), its content is first written into a variable and then served to the user. Such behavior can use up the whole memory of the system and crash the server process. 

\subsection{Pipeline Design}\label{sec:pipeline-design}
To investigate the spread of this particular vulnerable code pattern across GitHub, an automated approach is necessary due to the vast number of candidate projects hosted on GitHub.
The primary design objective of this approach is that it needs to select candidate repositories, detect a vulnerable code snippet in such a repository, attempt exploitation, and finally remediation, all in an automated fashion with as little manual effort as possible.
A pipeline is a natural design for such an approach, particularly because every step of it also acts as a filter: at any step in the pipeline, a candidate project may be discarded. Our pipeline components and their interaction are shown in Fig.~\ref{fig:flow}.
Steps A, B, and C automate the detection of potentially vulnerable open-source projects by sending search queries to the GitHub code search engine. Steps D and E attempt to automatically run and exploit a potentially vulnerable project. Finally, if the project is found to be exploitable, steps F and G automate the remediation of the vulnerability by creating a patch using an LLM, calculating the CVSS score, and submitting a report to the project maintainers.


\subsection{Entry Point (Step A)}\label{sec:entrypoint}
The first step is not necessarily automated, but a mandatory manual step to bootstrap the pipeline that only has to be done once for a vulnerability.
At this point, a vulnerable code pattern for a study has been chosen (as we do in Section~\ref{sec:snippet}).
To be able to detect a wide variety of instances of this vulnerable code pattern in GitHub repositories, one needs to search on the basis of keywords rather than a particular instance of the code pattern.
Within this step, one therefore has to manually extract specific keywords from the selected code snippet, preferring those that are less likely to change.
For example, a function name from a library is a better keyword compared to the name of a user-defined function or variable.
For the path traversal code pattern studied in this paper, we chose $\langle$\texttt{http.createServer, fs, read, URL, path}$\rangle$ as the primary keywords.




\subsection{Retrieving Candidate Repositories (Step B)}\label{sec:search}
With the vulnerable code pattern keywords, we search for open-source projects that contain such keywords. To do this, a code search engine is required, which has already indexed open-source projects or can crawl the projects and search for the code. 
Such an engine's quality and project coverage highly affect the quantity of the results from the pipeline.

We focus on GitHub as the most prevalent code hosting platform, which allows vulnerabilities to be assessed in realistic deployment environments. 
An alternative to using GitHub could be to work with a dataset used for studying open-source software. Such datasets usually consist of projects mined from GitHub. We found that such datasets are outdated or often contain very few projects compared to the existing projects on GitHub~\cite{gousios2013ghtorent,gharchive}.
For example, as of April 2025, there are over 254 million public repositories on GitHub~\cite{githubSearchPublicRepos}\footnote{We have seen an unstable result count from the GitHub search feature}.
At the same time, as of April 2025, the GitHub Activity dataset on Google BigQuery (last updated in May 2016) contained code from 2.8 million repositories, but only 400,000 projects were collected completely (i.e., only 0.002\% of public repositories available on GitHub). Thus, in our study, we work with the GitHub Code Search API as the search engine.

A major issue with this API is that the results are limited to 1,000 (10 pages with a page size of 100) samples only~\cite{romano2021g,dabic2021sampling}. Previous studies surpassed this limitation by adding extra criteria in GitHub Search for repositories. However, GitHub Code Search does not allow such criteria to be included in the query. To the best of our knowledge, the current search results limit (1,000 samples) in GitHub Code Search API is an open issue~\cite{githubDiscussion9868} (an answer from GitHub staff states that they intend to eventually improve the API to output the whole search results). 

To overcome this 1,000-result limit, we propose a recursive algorithm that refines the search based on newly discovered keywords.
We use a feature described in the GitHub documentation: if a new keyword is added to a previous set of keywords, the final files in the results will contain all of the keywords.
This allows us to reduce the search space and retrieve more projects by adding less frequently used keywords.
Our recursive algorithm first tries to get 1,000 results using the base query extracted from the vulnerable code pattern (the primary keywords listed in Section~\ref{sec:entrypoint}). Then, the query will find all unique words in downloaded code files and sort the words based on their probability of existence in the extracted files, using the tf-idf measure. For example, terms used to declare variables such as \texttt{const}, \texttt{var}, \texttt{let}, and library inclusion such as \texttt{require} have very high tf-idf scores, as they are the most likely to be included in all files. Such frequent keywords will not help limit the results. At the same time, to avoid wasting the API queries, we also aim to exclude very rare keywords that will only match very few results. Thus, we choose keywords that have at least 100 results. When the results are above 1,000 samples (the maximum number of possible samples returned by the GitHub search API), a new keyword from the previous set will be chosen to limit the search space. 
Finally, to prevent duplicates, we store links to the latest version of the searched files in a database.

\subsection{Static App Security Testing (Step C)}\label{sec:sast}
Because the search engine is queried with keywords rather than a specific vulnerable code pattern in the previous step, the results may include false positives.
These are projects that contain the keywords but not the specific vulnerable code pattern.
Before moving on to the dynamic exploitation of candidate projects, we prune the search results using less computationally expensive static analysis techniques.
Specifically, we use the popular static analyzer SemGrep~\cite{semgrepwebsite} for taint analysis with publicly available rules. Note that the quality of the rules highly affects the final results, and this step may not eliminate all false positives~\cite{bennett2024semgrep}. 
One reason is that developers do not always come up with the same solution to fix a vulnerability; thus, understanding and applying all the fixes in taint analysis rules is a tedious task with many corner cases~\cite{bandara2020fix}. 
False positives that remain after this step (projects identified as vulnerable but actually secure) will be eliminated in the next steps of the pipeline.

\subsection{Runner (Step D)}\label{sec:runner}
Candidate projects that reach this step are labeled as vulnerable by static analysis, but it has not yet been confirmed whether the vulnerability can be exploited (or is actually present).
To confirm exploitation, the first step is to automatically determine how to run the project and whether the launched process satisfies the prerequisites for exploitation.
For the Node.js projects in our case, we first try to automatically execute the collected vulnerable files in a Docker container. Three methods were implemented to automatically run the vulnerable Node.js file: (1) run the file by directly interpreting the code with the \texttt{node} command; (2) first install the project dependencies and then run the \texttt{node} command; (3) install dependencies and execute \texttt{npm run start}, which usually results in running the script that would run the project. The order of these methods is chosen to speed up the pipeline based on the time required for each method, from fastest to slowest. After the execution is completed, we validate the prerequisites for exploitation of the path traversal vulnerability: we try to detect if the \texttt{node} process is still running, and if it is running, we detect if there is any open port assigned to the process, as the code is expected to create a static file server. The sample will not move to the next step if no open port related to the program is detected.

\subsection{Exploiter (Step E)}\label{sec:exploiter}
The automatic verification of the exploit performed by this step eliminates all false positives received from the previous steps.
This step requires specific exploit implementations that must be prepared separately for each selected vulnerable code pattern (step A).
In our case, the chosen vulnerable code pattern accepts the input in the path part of the URL, not in a parameter. Thus, this step is implemented using a list of payloads for path traversal attacks and their bypasses, such as double-encoded URLs, nested path bypasses, different starting points, etc. 

As we mentioned, the studied pattern has two possible exploit use cases: (1) path traversal and (2) denial of service. Moreover, the attack vector of the vulnerability can be \emph{local} or \emph{network}, based on the network interface that the program uses, which is further discussed in Section~\ref{sec:cvss}. We now discuss our automated exploitation of the two cases.

\subsubsection{Path Traversal Exploit}
The chosen vulnerable code pattern comes from the pathname part of a URL; therefore, it is not assigned to a specific query string. This allows skipping taint analysis and the exploit to be present at the pathname part. To verify the successful exploitation of a path traversal attack, a text file with random hexadecimal content is placed in the root directory of the Docker container named \texttt{/flag.txt}. Then, a series of HTTP \texttt{GET} requests is sent to the open port of the program. All of these requests try to read the \texttt{flag.txt} file using common payloads and bypasses for path traversal attacks. For example, we try \texttt{GET http://localhost:port/../../../../../} \texttt{flag.txt}. If the HTTP response contains the content of the flag file, and the web application itself is not placed in the root directory of the container, we consider the program vulnerable to path traversal, as a file was served outside of the intended directory.

It is worth mentioning that HTTP clients, such as web browsers or \texttt{curl}, mostly normalize the pathname part of the URL by default. This will prevent exploitation, as the previous example of exploitation will be converted to \texttt{http://localhost:port/flag.txt} before it is received by the server. We believe that this client-side normalization might be one of the reasons developers think that the code is secure. However, attackers are not restricted to using such well-behaved HTTP clients. This will be discussed in Section~\ref{sec:rootcauses}.

\subsubsection{Denial of Service Exploit}
Sometimes the code pattern loads the entire content of the desired files in memory first and then sends it back to the client. This pattern creates a denial-of-service vulnerability, which leads to high memory usage and crashes the process. To identify which projects are vulnerable to such an attack, we limit the running memory usage of the Docker container to 1GB and bind the \texttt{/dev/urandom} file as \texttt{/flag.txt}. Then we follow the same approach in the path traversal exploit (send HTTP requests to read the \texttt{flag.txt} file). If the \texttt{node} process exits within 10 seconds of running the exploit, it means the file's content used the whole available memory and crashed the process; thus, the code is also vulnerable to a denial-of-service attack.


\subsection{Patcher (Step F)}\label{sec:llm-patcher}
Although the same vulnerable code pattern is shared across projects, the actual implementation differs per project due to variations in variable naming and code styles. Thus, patching all these projects requires tailored solutions, and it is not feasible to do it manually. Therefore, we apply large-language models (LLMs) to create patches for the different projects. Note that since all projects are already publicly available on GitHub, submitting small code snippets from these projects to an LLM does not present a confidentiality issue. 

\subsubsection{Patching Considerations}
Our goal is to produce small and consistent patches (i.e., cross-platform) that will not have side effects on the functionality and security of the code (besides fixing the intended vulnerability). If the proposed patch is small, it will be easier to understand for the maintainers, who might prefer to verify the patches themselves, instead of relying on fully automated solutions~\cite{winter2022developers} -- and, of course, in our setup, they have no reason to trust patches proposed by an unknown third party. Furthermore, the path traversal vulnerability appears to be a challenging target for state-of-the-art automated program repair tools, with the majority of them unable to provide fully correct patches~\cite{bui2024apr4vul}.

We observed that simply prompting an LLM chatbot to fix the path traversal vulnerability results in too many random outputs, such as modifications to other parts of the code, poor code quality, or new security vulnerabilities being introduced. For example, during our tests, we encountered patches that showed the requested path in the content of the HTTP response. Such behavior can introduce reflected cross-site scripting vulnerability (XSS) when the content-type is \texttt{text/html}. 

To address the issues with LLM-designed patches and keep the patch consistent and cross-platform, the first and second authors, each with over 5 years of experience in the web security field, designed a mitigation method. There are several methods to mitigate this path traversal vulnerability. For example, there are operating system layer approaches (such as chroot, containers, AppArmor, etc.), and application layer mitigations (such as comparing the intended public directory with the requested path, normalizing the pathname, removing double dots, etc.). As the vulnerability is contained in the application-level code, we aim to patch it at the same level, keeping the patch simple and local. 

At the same time, the detection of the intended public directory can be challenging, especially without a full context of the project, and would increase the complexity of the patch. Removing double dots would potentially also affect existing variables and may conflict with other existing code, resulting in a greater risk of side effects. Another solution is to drop the request when the normalized path is not equal to the requested path, which signifies that the request contains suspicious characters. However, a compatibility issue remains: while this approach works on UNIX-like systems, on Windows, the \texttt{path.normalize} function replaces slashes with backslashes, causing even benign requests to be dropped. Based on these considerations, we chose an approach to simply drop the requests that contain double dots. This approach does not modify other variables, works on different operating systems, is secure, and is simple to apply and review. Finally, if applying this mitigation fails, this will be detected in our patch validation step. For consistency, we did not include the path separator character; however, it can be included to minimize corner cases, e.g., if a file contains two dots in the filename.

\subsubsection{Generating Patches Using LLMs}
\begin{figure}[t]
  \centering
  \includegraphics[width=0.76\linewidth, keepaspectratio]{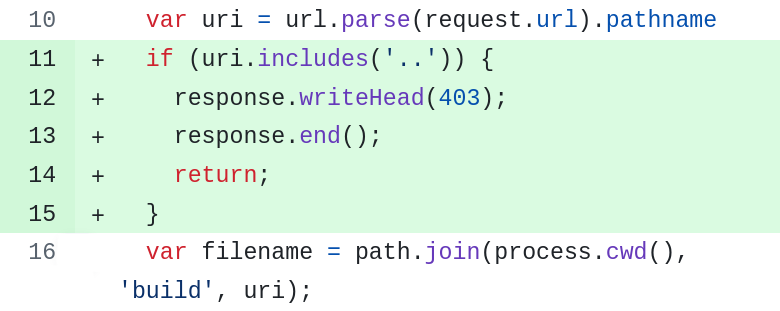}
  \caption{Sample patch (in green) generated using GPT-4 with our prompt.}
   \label{fig:samplepatch}
\end{figure}

The prompt instructs the chatbot to generate the patch to return as early as possible and to treat the URL similarly to the way it is used in the code. Additionally, the prompt instructs the LLM to perform no variable assignment and only block the incoming request. A sample of our custom prompt to generate a patch file can be found with the pipeline code\footnote{\url{https://github.com/JafarAkhondali/DotDotDefender}}. Fig.~\ref{fig:samplepatch} shows an example patch generated for an open-source project with such a prompt. It is worth mentioning that patches are generated completely by an LLM. Thus, it is straightforward to change it further for other patching methods or different vulnerabilities. 

To select a chatbot service for this work, the patching process was tested on a few samples using several chatbot services (ChatGPT, Gemini, Copilot, Claude). The first and second authors manually validated the quality of the patches for these samples. Based on this validation, we chose GPT-4 from OpenAI to patch the vulnerability. Recent literature supports our empirical evidence that GPT-4 is the better-performing LLM for vulnerability patching~\cite{sagodi2024,nong2024}.

\subsubsection{Patch Validation}\label{sec:patch-test}

To minimize the side effects, the prompt to create a patch is designed not to modify existing variables in the source code and to only block malicious input. However, as LLMs are not always predictable, we defined several rules to verify that the generated patch works correctly. Note that the vulnerable samples patched in this step have already been proven to be vulnerable and exploitable in a previous step. Thus, it is not required to exploit them again. Verifying the quality of the patch requires us to ensure that: (1) the patch does not break the application; (2) the patch does not change the functionality of the code; and (3) the patch correctly blocks malicious requests. To achieve that, before applying the patch, we send a simple benign HTTP \texttt{GET} request to the root address of the web server (e.g., \texttt{http://127.0.0.1/}) and store the HTTP status code as the expected result. Then, after applying the patch, we verify the following conditions: (1) the method used to run the project is not changed (e.g., if the vulnerable file was runnable using the \texttt{node} command, then running the project with \texttt{npm start} is not acceptable as it means the previous method failed and application is broken); (2) when sending a benign \texttt{GET} request to the web server root, the HTTP status code is not changed; (3) none of the exploits can read the file in the file system root (e.g., \texttt{http://127.0.0.1/../../../etc/passwd} should not work); (4) the sanitizer that blocks requests where the pathname in the URL contains two dots was placed correctly. 

The fourth check is implemented using a custom CodeQL (GitHub production-grade tool for static analysis)~\cite{ghcodeqlreference}\footnote{CodeQL was used in this step as the free version of SemGrep does not provide the necessary features.}. This rule compares how many times the \texttt{includes} method was applied to a pathname before and after applying the patch. If this number is higher after applying the patch, this is considered a sign that the sanitizer (as in Fig.~\ref{fig:samplepatch}) was added by the patch produced by the LLM and that the pathname is now correctly sanitized. The reason we counted the number of \texttt{includes} method invocations using the CodeQL rule on both the vulnerable (before) and patched code is that in some cases, the code may already contain the \texttt{includes} function for other purposes.

Overall, our first and second criteria contribute to the verification of the functionality of the app, and the third and fourth criteria contribute to the verification of the correctness and robustness of the patch. Nevertheless, due to the complexity and variability of the projects, it is not possible to fully guarantee the correctness and robustness of the LLM-generated patches. We discuss this further in Section~\ref{sec:discussion}.

\subsubsection{Prompting for Patching}
In the prompt, we ask the chatbot to output a patch file instead of the whole file content to reduce the token usage and limit non-security-related changes between input and output. Patch files are required to contain accurate line numbers for changes. However, we observed in an experiment that responses from the chatbot were not accurate in detecting line numbers. To counter this, we changed the prompt to include line numbers together with the source code as shown in Fig.~\ref{fig:samplepatch} in the appendix.
This resulted in a much more stable response. To apply the patch seamlessly, we also used some flexible options to allow more robust patch line number detection and correction on applying the patches.

Even with our extensive prompt and the above countermeasure for patch accuracy, the responses from LLMs are not consistent. That is, the responses are not identical per prompt. While this can be seen as an issue, making the responses consistent is also not very useful for our use case: if the prompt is tuned to be 100\% reproducible, we will always get the same response for a patch. In case this patch is corrupted (e.g., bad format or insecure) the LLM will be unable to correct the mistakes. For the final pipeline version, we chose $1.2$ as the temperature parameter and used GPT-4 as the LLM model. Suppose any of the four criteria for patch verification fail for a project after patching. In that case, the pipeline will retry the prompt with the same configuration until the three criteria are satisfied or the maximum number of retries is reached (by default, 20 times).

\subsection{Reporter (Step G)}
The next step in our pipeline is to estimate the severity of the confirmed vulnerability and to responsibly disclose the vulnerability to the project maintainers. 

\subsubsection{Calculating CVSS Score}\label{sec:cvss}
CVSS (Common Vulnerability Scoring System) is the standard metric used to assess the severity of security vulnerabilities. The CVSS version 3.0~\cite{cvssv3spec} combines several metrics named Attack Vector (AV), Attack Complexity (AC), Privileges Required (PR), User Interaction (UI), Scope (S), Confidentiality (C), Integrity (I), and Availability (A). Calculating the CVSS scores precisely for each vulnerability is a challenging task and requires estimates by security experts. In our case, the values of some of the mentioned metrics can be set the same for all of the discovered vulnerabilities, and some of them are different per target. We now explain how we compute the value of the metrics, based on the definitions in CVSS version 3. 

\subsubsection{Metrics Identical for All Projects}
\textbf{Attack Complexity} is set to \textbf{low}, as there are no specific configuration requirements for exploitation. \textbf{Privileges Required} is set to \textbf{None} as the Exploiter only sends an unauthenticated HTTP request. \textbf{User Interaction} is set to \textbf{None} for the same reason. \textbf{Scope} is set to \textbf{Unchanged} as there is no specific scope-changing methodology that would apply to all projects. Note, however, that if it is possible to change the scope, then the impact would be higher; thus, our CVSS estimate is conservative. \textbf{Confidentiality} is set to \textbf{High} as it is possible to read any file on the server with the privileges of the web server. \textbf{Integrity} is set to \textbf{None}, as there was no direct approach to change data with this vulnerability.

\subsubsection{Project-Dependent Metrics}
The remaining metrics can have different values per project, and we implemented dedicated methods to calculate them. \textbf{Attack Vector:} In the context of this vulnerability, the attack vector can be either \textbf{Network} or \textbf{Local}. To identify the accurate value for each vulnerable program, we first run the exploitation on the remote address of the network using the Docker container. If the exploit is successful, we assign \textbf{Network}, otherwise, we try exploiting it locally by executing the Exploiter inside the Docker container and using \texttt{localhost} as the hostname to verify the vulnerability exploitation at the local address. Note that if none of these methods work, we will not consider this project to be exploitable. \textbf{Availability} impact is a potential side effect of the detected vulnerability. As tested in the 
Exploiter (Step E), the value of the Availability metric in CVSS calculation is set to \textbf{High} if it is vulnerable and otherwise, it is set to \textbf{None}.

\subsubsection{Vulnerability Disclosure}\label{sec:disclosure}
Responsible disclosure of the discovered vulnerabilities is one of the most challenging parts of the pipeline. Different companies and projects prefer different methods to receive vulnerability reports. Responsible disclosure requires private communication with maintainers: it is a time-consuming process if done for a large number of projects. At the same time, many projects are not maintained or are maintained by security-unaware developers who might not understand the vulnerability or the patch shared with the disclosure. 

We have developed a staged and diversified responsible disclosure approach to balance the benefits of notifying as many maintainers as possible against the potential harms of a massive vulnerability disclosure via GitHub. First, we randomly opened issues in several projects with under 200 stars. The issue did not contain any sensitive details and only asked to enable the GitHub private security feature, so that we can communicate privately, or propose another place to discuss the vulnerability. Then, if maintainers were interested, we gave them the full disclosure. With some feedback from maintainers, we decided to focus on projects with more than 5 stars, as some projects are not used in production. In the next stage, we sent pull requests to several projects with more than 5 stars and waited for feedback. Some maintainers never responded or replied that the projects were not maintained anymore; thus, finally, we have decided to limit the disclosure (opening an issue followed by a pull request) to projects with more than 5 stars that are maintained (latest commit within 365 days). 

For ethical reasons, we decided against opening issues on popular GitHub projects (200+ stars), as threat actors can potentially see which repositories we opened issues on and guess the vulnerability. Due to the popularity of such projects, their many users might be affected. Thus, we reported vulnerabilities to such projects manually, to e-mail addresses associated with the projects (if available). We also attempted to find relevant contact points by searching for security reporting policies. For this, we automated 
the search and parsing of security policies. Another option was to send direct e-mails to developers by scraping their e-mail addresses from the committer section of their commits, but we chose not to do so: doing it on a large scale might compromise the privacy of developers. Moreover, such e-mails could end up in spam folders. 


\subsubsection{Pull Request Submission}
We monitor the status of all vulnerable projects. If the exploitable vulnerability is still present after 30 days of our report, we will also send a pull request with the fix automatically. We implemented this step by forking the repository, applying the previously crafted patch, and sending the changes as a pull request. We decided to send pull requests even in projects where the maintainers considered the impact to be low. The reasons we do this are: (1) accepting or rejecting a pull request (especially without side effects to other variables) is trivial; and (2) some developers mentioned to us they would be happy to receive a pull request for the fix, but they could not spend time on fixing the problem themselves.

After sending vulnerability reports, we asked developers to participate in a short, optional survey to gather feedback about the quality of the reports, their satisfaction, and opinions about the vulnerability. The feedback received from the maintainers is discussed in Section~\ref{sec:survey}.

\section{Prevalence of the Vulnerable Code Pattern in GitHub}\label{sec:statistics}

To answer our question about how many more open-source projects out there contain the vulnerable path traversal code pattern, we initialized our pipeline with this code pattern. We executed the pipeline against all public GitHub projects as the data source. In this section, we report on the results of executing our pipeline. We summarize the popularity of the vulnerable projects, the CVSS score of the confirmed vulnerability instances (with the median being 9.1), and a number of other relevant statistics in Appendix~\ref{sec:more_details}.



Table~\ref{tab:results} presents the summary of vulnerability statistics from each step of the pipeline. In total, our search method resulted in 40,546 open-source projects scraped from GitHub. In these projects, 41,870 unique files contained the vulnerable keywords under 33,364 unique GitHub usernames or organizations. The pipeline failed to download 37 of these results. Most of the results were labeled as secure by the static analysis step, thus, SemGrep identified 8,397 samples as vulnerable and 33,436 samples as safe. The Runner failed to automatically run 3,908 samples using the defined methods (we examine some reasons behind this in Section~\ref{sec:pipelineissues}). Out of 4,489 candidates for automatic exploitation, 1,756 were successfully exploited and 2,733 were not exploited. Among 1,756 vulnerable files, 484 samples were also vulnerable to denial of service (as discussed in Section~\ref{sec:cvss}). A list of successful payloads is presented in Table~\ref{tab:pocsstats} in Appendix~\ref{sec:more_details}. In the Patcher step, 1,600 valid patches were generated and validated automatically (as explained in Section~\ref{sec:patch-test}). For 188 samples (out of 1,756), the produced patches could not be proven with our tests not to break the software. Out of these 188 samples, there was only one popular project (177 stars). The report for this project was sent manually, and it was fixed by the maintainers.

We have approached 31 very popular projects manually, and 13 of them have been patched (remediation rate of 42\%). We created 245 issues at prominent projects with fewer than 200 stars, of which 70 projects reacted. We have sent 433 pull requests, with 46 of them being accepted and 24 being closed. Additionally, 4 projects were manually patched by the maintainers (without accepting our patch). Thus, the total remediation rate among projects that received a pull request is 11\%. In total, 63 out of 464 reports fixed the vulnerability (remediation rate of 14\% among the projects that received full vulnerability and patch information).


\begin{table}[t!]
  \centering
  \caption{Summary of vulnerabilities.}
  \label{tab:results}
  \scalebox{.9}{
  \begin{tabular}{>{\raggedright\arraybackslash}p{7cm}r}
    \toprule
    \textbf{Metric} & \textbf{Count} \\
    \midrule
    Total projects analyzed (step B) & 40,546 \\
    Unique files with vulnerabilities (step B) & 41,870 \\
    SAST vulnerable samples (step C) & 8,397 \\
    SAST safe samples (step C) & 33,436 \\
    Automatic running failed (step D) & 3,908\\ 
    Successful exploitation (step E) & 1,756 \\
    Generated valid patches (step F) & 1,600 \\
    \midrule
    Pull requests sent to projects (step G) & 433 \\
    Pull requests accepted & 46 \\
    Manual fixes by maintainers (based on public pull request) & 4 \\
    Pull requests closed & 24 \\
    Private report made to maintainers & 31 \\
    Manual fixes by maintainers (private report) & 13 \\
    \midrule
    Total fixed repositories & 63 \\
    \bottomrule
  \end{tabular}
  }
\end{table}


\section{Developer Feedback}\label{sec:survey}
To collect feedback from developers and evaluate how they assess our work, we asked maintainers to participate in a short, optional survey. 
The survey questions are listed in Appendix~\ref{app:survey}.
We received 15 responses from the survey. 
Developers mostly strongly agreed that such reports help improve security.
We specifically asked the maintainers to rate the quality of each part in the final generated report they received. 
The maintainers responded mainly with positive feedback, and  
the majority of developers who replied believed that the code was vulnerable and not a false positive. 

Note that the results are only from the developers who participated in the survey, not the actual number of fixes. The statistics on how many projects were fixed after our report are discussed in Section~\ref{sec:statistics}. In general, only a small fraction of project representatives participated in the survey, which is, unfortunately, common in survey research with developers~\cite{smith2013improving,kaur2022recruit}.

In the set of rejected pull requests (24 cases), two maintainers mentioned the code was intended for development purposes only and was not deployed in production. One maintainer noted that the vulnerable web server was used for testing only; another explained that the project was unmaintained. In four cases, the maintainer chose to implement their own patch instead of accepting the pull request. One of the repositories of these four cases was related to a web development course at a top US university. The maintainer verified the issue and implemented the fix into a newer version of the course material. Finally, two maintainers assumed that the report was spam, and one mentioned that the vulnerable part was not used in their software. From these responses, we can conclude that for some affected projects, the potential impact is estimated by the maintainers as low, and there is sometimes a lack of trust or interest from the maintainers who might ignore a valid vulnerability report. This is in line with a recent study on vulnerability management in open source projects~\cite{ayala2024mixed}.

\section{Root Causes of the Vulnerability}\label{sec:rootcauses}

Based on the vulnerable code patterns, we tried to identify the sources of these code patterns and to discover why this vulnerable code pattern propagated in so many projects and did not receive a fix earlier. In many cases, this vulnerable code pattern is present for the whole lifecycle of the software project. It is important to note that we want to study the root causes to learn some useful lessons and prevent similar issues in the future, not to put the blame on people or organizations.

\textbf{The sources of the vulnerable code pattern.}
The oldest occurrence of the vulnerable code snippet we found in the GitHub Gist~\cite{oldestvulnseen} was created in 2010. GitHub Gist is a GitHub service for sharing code snippets, and previous research has shown that many shared gists have security issues~\cite{rahman2019share}. Gists are usually less popular and known compared to repositories; however, this Gist has 543 stars and 204 forks, which is quite uncommon. Interestingly, in 2012, a developer commented that the code was vulnerable. Then, in 2014, another developer raised the same concern about the vulnerability, and yet another developer mentioned that the code is safe, after testing. In 2018, somebody commented about the vulnerability again, and another developer insisted that that person did not understand the issue and that the code was safe. 

The next popular place of this code snippet was a hard copy of a document created by the community in Mozilla developers in 2015 and was fixed in 2022~\cite{mdndocsfirstfixed}. However, the vulnerable version also migrated to StackOverflow in late 2015. Although it received several updates, the vulnerability was not fixed: the code snippet there is still vulnerable~\cite{sovulnanswer}. The same story happened in 2016 in another StackOverflow question (with over 88,000 views), where the same pattern was used, and a developer suspected the vulnerability. Yet, as this developer was not able to verify the issue, the code was assumed safe~\cite{sovulnquestion2}. Similarly to what happened with Gist, a user raised the concern of vulnerability, but no one provided a patch or a proof-of-concept (PoC) exploit, leaving the vulnerable code as it is. We note that the StackOverflow question page has been viewed over 758,000 times; this is the $115$th most viewed question in StackOverflow for the \texttt{Node.js} tag~\cite{somostquery}. We have also found several Node.js courses that used this vulnerable code snippet for teaching purposes~\cite{nodecourse1,nodecourse2,nodecourse3,nodecourse4}.

\textbf{Why developers assume it is safe.}
As mentioned before, we believe the main reason for this misunderstanding between developers, which lasted for over 15 years, is that when developers test the code, they usually use a web browser or the \texttt{curl} command. These clients will normalize the URL by default. For example, to exploit this vulnerability using \texttt{curl}, the tester must use the \texttt{--path-as-is} flag, which means \textit{Do not squash .. sequences in URL path} according to the \texttt{curl} documentation. By simply providing a working proof-of-concept exploit or editing the question, it would be possible to overcome such issues. We, again, note that the attackers are not bound to use the standard clients, and, thus, the vulnerability is exploitable in the wild.

\begin{figure}[t]
  \centering
  \includegraphics[width=0.86\linewidth, keepaspectratio]{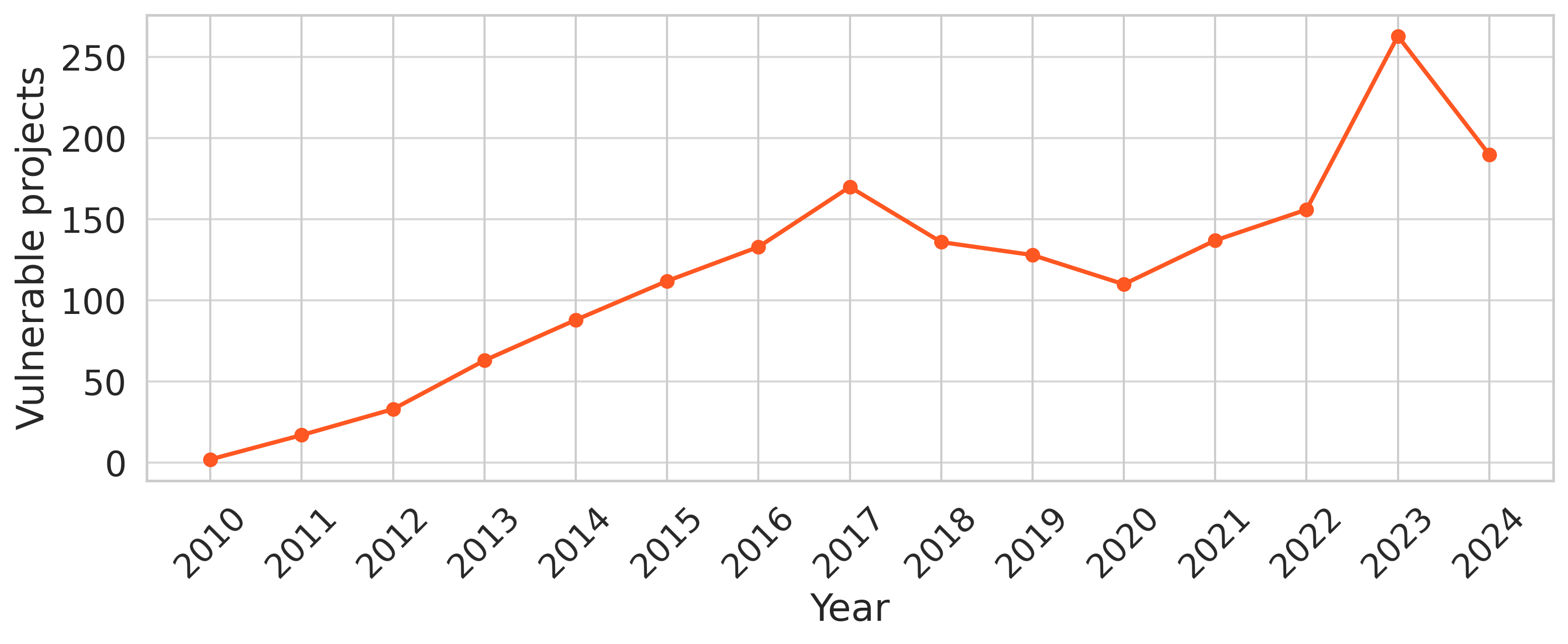}
  \caption{Distribution of verified exploited samples per year.}
   \label{fig:commitdateyears}
\end{figure}

The approximate commit date of the vulnerable code in the exploited repositories is shown in Fig.~\ref{fig:commitdateyears}. It shows a growth trend up to 2017, which is close to the date of the discussed sources of vulnerable code. Another growth trend can be seen from 2019, which might be due to one of the free popular Node.js courses released in 2019; however, more evidence is required for such a hypothesis. It can also be expected that, given the overall rising number of developers and projects, the number of vulnerabilities would rise as well. We stress that it is very important to raise awareness about this (and other) vulnerabilities in the developer community. This is why it is critical to responsibly communicate the information about the vulnerability and the patch to the maintainers. To raise awareness, we have provided information about this vulnerability at the original GitHub gist and the two StackOverflow questions.

\section{Side-Effects on LLMs}\label{sec:llmseffects}
Large-language models are trained on large-scale datasets, including publicly available projects on GitHub, StackOverflow discussions, and other internet sources. The studied vulnerable code pattern appeared many times, with different styles. To understand the effect of these vulnerable code snippets on LLMs, we experimented with several popular chatbots. We designed 2 scenarios: (1) prompt the LLM to create a static file server without third-party libraries, and then ask in the same conversation to make it secure; (2) prompt the chatbot to create a secure static file server without third-party libraries. The exact prompts we used in this experiment are available in the supplementary material.

\begin{figure*}[ht]
  \centering
  \includegraphics[width=0.76\textwidth, keepaspectratio]{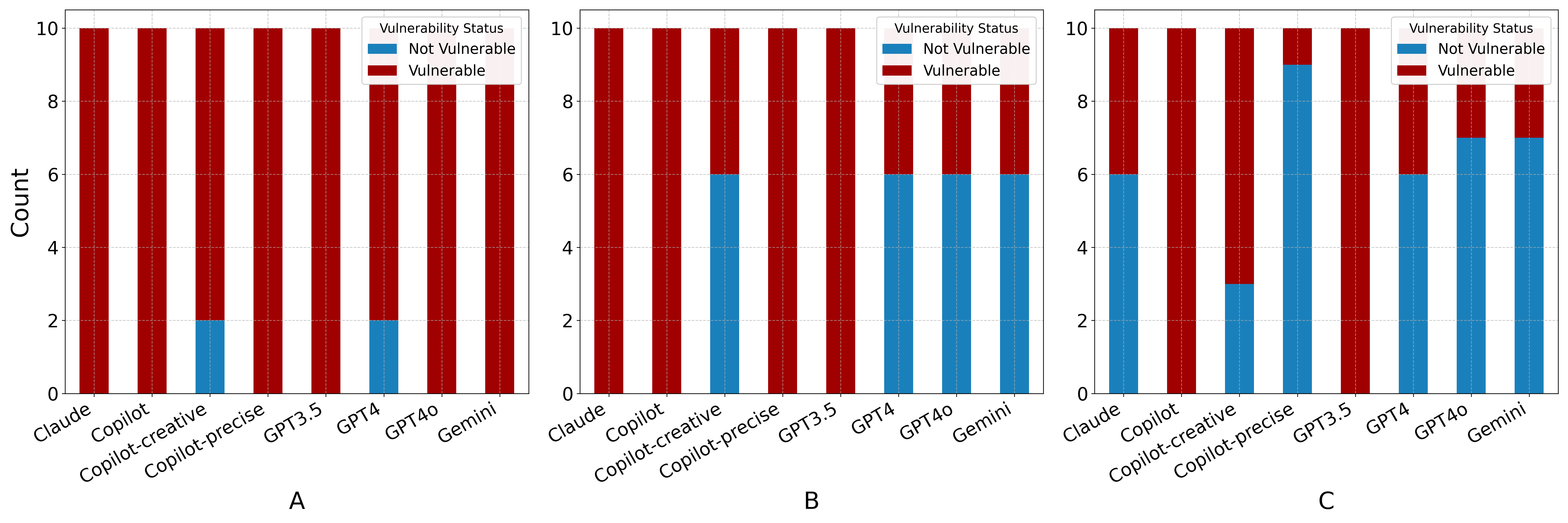}
  \caption{Distribution of vulnerable code snippets generated by LLMs in different scenarios: (A) scenario 1, step 1; (B) scenario 1, step 2; (C) scenario 2.}
   \label{fig:llmsmixed}
\end{figure*}

For each scenario, we repeated each procedure 10 times per LLM. After prompting, we collected all the produced code snippets and tried to exploit them. As can be seen in Fig.~\ref{fig:llmsmixed} part A, in the first scenario (without mentioning security), most of the LLMs generated vulnerable code to path traversal attacks (76 vulnerable out of 80; 95\% vulnerable). In the second step of the first scenario, as shown in part B, by asking LLMs to secure the code, the results improved (42 vulnerable out of 80; 52.5\% vulnerable), still showing that the majority of the samples were vulnerable to the attack. In the second scenario, where we prompted the LLMs to generate secure code (part C), the ratio of the vulnerable samples is very considerable (56 vulnerable out of 80; 70\% vulnerable). We note that GPT-3.5 and Copilot (balanced) did not generate secure code in any scenario. The only model that performed better in the second scenario compared to the first scenario was Copilot-creative. Copilot-precise achieved the highest score when asked to make the code secure if it was not.

Previously, it was shown that developers copy and paste vulnerable code on code-sharing platforms such as Stack Overflow, and such code snippets are then migrated to GitHub projects~\cite{fischer2017stack,verdi2020empirical,chen2024empirical}. Nowadays, with the advances of generative AI, the popularity of StackOverflow and similar platforms is decreasing in favor of LLM chatbots~\cite{da2024chatgpt}, which also learn from open community resources~\cite{hamer2024just}. A big advantage of a community-driven code-sharing platform is the ability to discuss code snippets, which facilitates a deeper understanding of code and security risks for developers (even though it does not always work perfectly, as we have observed). Lacking such a feature in the LLM chatbot thus increases security risks, as the only source to rely on is the chatbot itself. 

This experiment shows that the popular LLM chatbots have learned the vulnerable code pattern and can confidently generate insecure code snippets, even if the user specifically prompts them for a secure version. Therefore, our study emphasizes that popular vulnerable code patterns need to be eradicated not only from open-source projects and developers' resources but also from LLMs, which can be a very challenging task.


\section{Discussion and Limitations}\label{sec:discussion}

\subsection{Summary and Implications of Our Results}\label{sec:summaryresults}

To identify the prevalence of the path traversal vulnerable code pattern across GitHub open-source projects (our objective \textbf{O1}), we designed and implemented an automated pipeline. To the best of our knowledge, this is the first pipeline in the literature that 1) includes all steps from candidate project finding to vulnerability discovery and confirmation and to security impact assessment through CVSS score calculation (thereby achieving our objective \textbf{O2}) patch design, and vulnerability disclosure to maintainers (addressing our objective \textbf{O3}), and 2) detects a new vulnerability (a zero-day for the impacted projects) rather than working on a dataset of known vulnerabilities. We outline the novelty further in Section~\ref{sec:relwork}. 

Our pipeline automatically detected path traversal vulnerabilities in 1,756 open-source projects on GitHub, and we were able to remediate at least 1,600 of them locally and 63 of them publicly.   

By examining the sources of this vulnerable code pattern that emerged throughout the past 10-15 years in various reputable developer communities and was largely approved as secure (objective \textbf{O4}), we have shown that developers need more support in understanding how to deal with this type of vulnerability and how to test for it correctly. Our remediation efforts so far reveal that the maintainers are not fully embracing our notifications (the remediation rate based on pull requests is 11\% and on manual notifications to more popular projects 42\%). This is worrying, and it is a signal that we need to develop more effective vulnerability communication and remediation channels. Especially today, when software supply chain attacks have become so prominent, efficiently patching open-source projects has become a critical issue. Additionally, we have shown the worrying impact of the proliferation of a single vulnerable code pattern on LLMs (objective \textbf{O5}). While it is not clear whether the biggest impact comes from the GitHub projects themselves or from developers' resources that distribute this code example, we think that a concerted effort from the community is needed to support developers in both securing their code and improving the resources they use.

\subsection{Limitations of the Pipeline Implementation}\label{sec:pipelineissues}
Our pipeline has certain limitations and, in the future, its capabilities can be extended.

\textbf{Search API Limitations.}
We designed our pipeline to work around the GitHub Code Search API limitation (1,000 results per query). However, our solution is not comprehensive enough to fully eliminate the search limitation. This limitation led to missing additional vulnerable projects, and therefore, the true prevalence of the vulnerability is higher. To the best of our knowledge, our proposed method is the only workaround that allows GitHub-wide code search. While GitHub code search was not the main focus of this work; however, improving it in the future will increase the repository coverage. A possible alternative is to use GitHub archives, but so far, we have not been able to identify a dataset that is sufficiently large and up-to-date. Moreover, developers often fork repositories without contributing to the original projects~\cite{githubforks}. For example, our dataset's most popular project (in terms of stars) in our dataset has over 5,200 forks.
Most of these forks are likely also vulnerable.  Some projects are simply a copy of another project that was pushed as a new project (shadow forks) and are also vulnerable. Such cases are orphan vulnerabilities and were previously studied in forks and file-level copies~\cite{reid2022extent}. We did not push patches to forks and did not investigate the popularity of vulnerable forks, as it was not a target of this research. However, if the fork owner simply updates the forks, the project would receive the patch.


\textbf{SAST.}
We chose a publicly available rule using SemGrep for taint analysis. The quality of such a rule and SemGrep itself can affect the pipeline. False positives (classified as vulnerable but not vulnerable) from this step are not an issue for the pipeline as they will be eliminated in the Exploiter step, but false negatives (classified as not vulnerable but vulnerable) can occur with the pipeline~\cite{bennett2024semgrep}. Moreover, removing the SAST filtering step and directly executing all projects could further reduce false negatives at the cost of increased runtime, and might reveal additional vulnerabilities, weak mitigations, or weaknesses in the SAST rules themselves. We manually checked a few samples (ca. 30) with high stars and observed no false negatives, but this step requires a deeper analysis. The focus of this work is to automate the process on a large scale with no false positives rather than focusing on false negatives.
In the future, multiple static application security testing tools can be used in the pipeline to make a joint decision~\cite{bennett2024semgrep}. 

\textbf{Runner.}
This module has some limitations related to software diversity. Some projects are too old and use functions already deprecated in Node.js. As the Node.js version of such projects was before version 1.0.0, we chose not to fix this issue in the Runner, as the projects were less likely to be used anywhere. In some cases, the port would not be open by default. For example, software that requires an OAuth flow for a callback login only activates the web server after receiving a specific command-line argument. Some programs would only accept a request once and then exit. Such cases require a more in-depth analysis of the code.
Moreover, there is no guarantee that all software will open a port. 


We selected the top 30 projects with the most stars that did not run automatically. After manually running them, 9 out of 30 samples were vulnerable. Thus, the Runner component limits the detection of vulnerable projects. However, even if the pipeline could not automatically run the project, it already helps in improving the efficiency of finding vulnerabilities by automatically collecting potentially vulnerable projects with keywords and filtering them using static analysis. We reported these vulnerabilities manually.

\textbf{Exploiter.} 
The vulnerable code pattern studied in this research used the pathname part of the URL directly without any special formatting, resulting in easy-to-implement exploit code. However, if the input was placed in a different place, such as a query string, our method would not work. This case requires a white-box fuzzing step or taint checking, which can improve the exploitation success rate; these can be added to future pipeline extensions. As this was not the case in the code pattern studied, we did not consider it. 

\textbf{Patcher.} 
Automated patching by LLMs, while drastically improving the scalability, has limitations as the patches might, in principle, introduce other bugs. We tried to minimize this risk by applying several checks mentioned in Section~\ref{sec:llm-patcher}. As we show in Appendix~\ref{sec:more_details}, the provided patches are relatively small. We also clearly explained in our communication with the maintainers that the patch has been generated automatically using LLMs. Ultimately, the responsibility to check the patch and correct it as needed lies with the maintainers, even if the patch would have been designed manually (or no patch was provided).  Additionally, due to the project's diversity, a small fraction of projects were not patched.

\textbf{Reporter.}
Some maintainers mentioned that the code was only used for development (not production), e.g., for testing. However, even if the vulnerable code is only for development, it can still make the system vulnerable to a network attacker. That is also the case with running test cases, as attackers still have an attack window. This makes a continuous integration server vulnerable while running test cases. Thus, our severity assessment is conservative. However, of course, we cannot fully understand the context of all vulnerable projects due to their large diversity.


There is no perfect way to securely contact the maintainers on a large scale. We used an active personal GitHub account of one of the authors with a good reputation on GitHub to help convince the maintainers that the issues were not spam. Still, some maintainers believed that we were spamming them. Unfortunately, opening hundreds of issues or pull requests in many GitHub projects under the same GitHub account is perceived as spamming and attracts negative reactions. Thus, we opted for a staged notification process. We got positive feedback after we publicly gave full vulnerability details to maintainers. However, doing this on a large scale offers malicious actors an attack time window until the fix is published~\cite{zhang2021survey}.

\textbf{Pipeline.} In our study, we strived to eliminate false positives, as reporting to maintainers a vulnerability that does not affect their code would be a major nuisance. Thus, we automated the collection of vulnerable projects, tested them using static analysis and by confirming exploitation, and then generated and validated a patch. The projects identified as vulnerable by confirming exploitation are vulnerable (no false positives). However, despite our validation attempts, the generated patches, as we mentioned, might not be fully correct, and the maintainers need to attentively analyze them before implementing. The staged pipeline design also results in false negatives: besides the GitHub Code Search API limitations, there might be, for example, many vulnerable projects among those that we could not run (as we mentioned in Section~\ref{sec:runner}). Thus, the overall incidence of the studied vulnerable code pattern on GitHub is likely much higher than what we report.  

Finally, the pipeline supports multiple vulnerable files per repository, but does not handle multiple vulnerabilities in a file. We found 36 repositories, each containing two vulnerable files.



\subsection{Extending the Study}
\label{sec:extending}

Whereas this study focuses on a particular code pattern, the pipeline design that we present is generic and can be adapted to various other vulnerability types.
To investigate other variations of path traversal vulnerabilities, the first step is to find other common vulnerable code patterns or other types of path traversal where the entry point is different.
Searching for common vulnerable code patterns can be done by mining real-world vulnerability datasets~\cite{bhandari2021cvefixes,akhoundali2024morefixes}. To find patterns, it is possible to combine such datasets in conjunction with code clone detection methods~\cite{sajnani2016sourcerercc,cordy2011nicad,lopes2017dejavu}.
Other variations of the path traversal or other vulnerabilities can also be found by using a SAST tool with customized rules.
Once a code pattern of interest is identified, it is trivial to bootstrap the pipeline with it (step A).
However, steps D and E would then require new tailored implementations to support confirming exploitation of the specific vulnerability exposed by the new code pattern.
These changes for another vulnerability only have to be done once, after which the pipeline will automatically search for and probe GitHub projects.

Moreover, the pipeline can be substantially enhanced by using LLM agents~\cite{zhang2025llms}. For example, to improve the Runner component, we can use LLM agents to run the projects automatically~\cite{hu2025llm, bouzenia2024you}. LLM agents are also promising in vulnerability exploitation~\cite{zhu2024teams,zhu2025cve,wu2024autopt,fang2024llmexploit}, vulnerability detection and false positive triage for SAST reports~\cite{10.1145/3664646.3664777, akuthota2023vulnerability, sheng2024lprotector,wang2023defecthunter}, and vulnerability management and repair~\cite{liu2024exploring,pearce2023examining,wu2023effective,xia2023automated,sagodi2024,nong2024}. The pipeline can also be improved by integrating more robust patch validation methods~\cite{fei2025patch,wang2020automated}, e.g., by running the test suites for projects equipped with them~\cite{bui2024apr4vul}.



\subsection{Ethical Considerations}\label{sec:ethics}
Following the Menlo report~\cite{bailey2012menlo,kenneally2012menlo}, we carefully balanced the benefits and risks to different parties when designing and executing our research. On the one hand, finding a vulnerability and proposing a patch to the maintainers improves the project's security and the security of the overall open-source software ecosystem. On the other hand, responsible vulnerability disclosure requires confidential communication between the reporter and the system owner~\cite{cavusoglu2005emerging}. This would mean that for all 1,756 projects, we need to establish confidential communication with the maintainers; however, this is not feasible given the scale. Moreover, it is important to limit the potential harm of large-scale vulnerability notifications via GitHub that can be observed by attackers, and are also perceived as spam by some maintainers. Thus, we opted for a diversified approach, whereby popular projects that are more likely to be in a supply chain for other users were contacted privately, preferably using the project-defined communication channels for security issues. For slightly less popular projects, we first create an issue and ask the maintainers to contact us for more details. After 30 days, if there is no action, we make a pull request with a patch and explanations about the vulnerability. Finally, we did not notify most of the unpopular or unmaintained projects. We acknowledge that the number of stars is an imperfect measure of project importance. 

Creating issues in many projects and submitting pull requests can get attackers' attention, and potentially endanger vulnerable projects from our dataset and those we missed. Yet, considering the important benefit of improving security~\cite{moura2023vulnerability}, we chose to report exploitable vulnerabilities in a way that would be acceptable to maintainers, which maximizes the chances of fixing the issue~\cite{ccetin2017make}. 


The response rate from the maintainers is relatively low (the total rate of patching among projects that have been notified is 14\%). While some maintainers communicated to us that the code is not in production and thus they do not consider the vulnerability to have security implications (which can be a fallacy, as we discuss in Section~\ref{sec:pipelineissues}), many projects are also not maintained anymore. Thus, it is important to consider the potential harm from attracting attention to this vulnerability, as many projects will remain unfixed~\cite{macnish2020ethics}. We believe that the benefit of raising awareness about the issue, especially among the very popular and influential projects and within developer communities, outweighs this potential harm.


Finally, we asked maintainers to fill in a survey. We did not ask for any personal data (i.e., those falling under the GDPR); therefore, the survey does not raise ethical issues related to personal data collection.  
A relevant Ethics Review Committee at the Science Faculty of Leiden University has reviewed and approved our study. 


\section{Related Work}\label{sec:relwork}
Automated vulnerability detection, exploitation, and patching are active areas of research. Until recently, \textbf{large-scale automated vulnerability detection methods} heavily focused on developing machine learning-based approaches that can learn to identify vulnerable code patterns from vulnerability datasets~\cite{russell2018automated,chakraborty2021deep,marjanov2022machine}. Other active research directions for automated vulnerability detection techniques are fuzzing~\cite{kim2020hfl,even2023grayc,mallissery2023demystify,zhu2022fuzzing} and leveraging patch information~\cite{li2022p1ovd,peng20191dvul}. We do not apply machine learning (ML) for detection, as we use a simple pattern matching and static taint analysis done by SemGrep~\cite{bennett2024semgrep}. We also do not apply fuzzing to detect vulnerabilities or generate exploits, because the vulnerable pattern is amenable to a more straightforward exploit design.  Closer to our work, \cite{kim2017vuddy,woo2022movery,woo2021v0finder,bowman2020vgraph,woo2023v1scan,feng2024fire} detect vulnerabilities similar to some already discovered ones by leveraging code similarity. Go et al.~\cite{go2023simple} use a static list of dangerous functions with GitHub Code Search to identify potentially vulnerable code. However, their approach lacks validation of exploitability. Leitschuh et al. leveraged GitHub CodeQL to perform a GitHub-wide search for vulnerabilities; however, handling false positives (e.g., by confirming exploitation) was not mentioned~\cite{jona22bh}. Moreover, GitHub removed the CodeQL search in 2022~\cite{lgtmservice}, which does not allow this methodology to be reused. 
Shcherbakov et al. designed a multi-label static taint analysis to identify prototype pollution vulnerabilities in
Node.js~\cite{shcherbakov2023silent}. By manually crafting exploits, they were able to exploit popular Node.js libraries (without providing patches). Kang et al. find JavaScript vulnerabilities using the abstract interpretation-based FAST method, which is also able to produce exploits that have to be applied manually~\cite{kang2023scaling}. In DAPP~\cite{li2021detecting}, researchers designed a static analyzer based on abstract syntax trees for detecting prototype pollution vulnerabilities in Node.js projects. Compared to our work, we focus on a different vulnerability, and instead of only static analysis, we automatically exploited candidate projects.

\textbf{Automated exploit generation approaches} can leverage, e.g., the fix commit and apply fuzzing or advanced code analysis techniques to automatically design and chain exploit gadgets~\cite{brumley2008automatic,avgerinos2014automatic,hu2015automatic,park2022fugio,bensalim2021talking,parameshwaran2015dexterjs}. For example, the taint analysis-based NodeMedic~\cite{cassel2023nodemedic} and fuzzing-based that is input structure aware, NodeMedic-Fine approaches~\cite{cassel2024nodemedic} automatically detect vulnerabilities in npm packages and generate exploits. Vulnerability repair methods often leverage ML methods and, most recently, transformer neural architectures to synthesize patches~\cite{fu2022vulrepair,shen2020survey,pinconschi2021comparative}. 


\textbf{Applications of LLMs.} Recently, LLMs have received significant attention as a means to detect~\cite{lu2024grace,zhou2024large}, exploit~\cite{xu2024autoattacker,fang2024llmexploit,fang2024llmhack}, and patch~\cite{chen2024large,kulsum2024case,pearce2023examining,liu2024prompt,jin2023inferfix,huang2024penheal,xia2023automated} diverse types of vulnerabilities and bugs~\cite{fan2023large}. 
For example, Pearce et al.~\cite{pearce2023examining}, Xia et al.~\cite{xia2023automated} and Wu et al.~\cite{wu2023effective} experimented with LLMs for automated vulnerability repair and found that the studied (earlier generations of) LLMs were not always able to correctly patch complex, real-world examples. More recent studies show that GPT-4 (used in our study for this task) performs better in vulnerability patching than other current LLMs~\cite{sagodi2024,nong2024}. 
Liu et al.~\cite{liu2024exploring} and Wu et al.~\cite{wu2024autopt} show that LLMs like ChatGPT are useful for penetration testing and vulnerability management tasks and can improve some of the individual steps of our pipeline. Fang et al. reported that LLM chatbots were able to exploit different CVEs~\cite{fang2024llmexploit} and hack insecure websites~\cite{fang2024llmhack}, while another study~\cite{fang2024large} shows that LLMs are capable of code analysis for security purposes. These works can deal with diverse vulnerability types, however, they do not include all of our pipeline stages, e.g., the vulnerability disclosure part.   
As we mentioned, future extensions of our pipeline can integrate LLMs to find vulnerable code patterns or produce exploits.  

Notably, LLMs do not always generate secure or even working code as shown by recent studies, e.g., \cite{yeticstiren2023evaluating,mousavi2024investigation,pearce2022asleep,khoury2023secure,liu2024your}. This is in line with our study results -- and we have shown that the insecure code pattern that we studied had likely contaminated the LLMs.

\textbf{Pipelines.} To execute our study, we developed a complete pipeline that not only scans GitHub and finds candidate vulnerable projects but also validates the vulnerability by exploiting it, generates the patch, and prepares a pull request for the maintainers. Very recently, there have been several efforts aiming at developing such pipelines, both in academia and industry. For example, Vulnhunter~\cite{vulnuntr2024} and Google's BigSleep~\cite{bigsleep2024} scan a given repository, identify vulnerabilities, and try to exploit them, thereby identifying zero-days in a given project. BigSleep, in particular, looks for variants of previously found and patched vulnerabilities. AutoPT~\cite{wu2024autopt} is capable of pentesting web applications, integrating a vulnerability scanner and a proof-of-concept exploit synthesis delivered by LLM agents. 


\textbf{Novelty of our work.} Compared to these studies,  our work introduces a complete, end-to-end pipeline that (1) identifies a real-world, widely used vulnerable pattern beyond simple function lists, (2) automatically executes and exploits the target code to filter out false positives, and (3) generates patches and submits pull requests with detailed vulnerability reports. To our knowledge, no prior work automates this full lifecycle from detection to mitigation and remediation on this scale, marking a significant advancement over existing methods. Furthermore, in our study, we examine the prevalence of a specific path traversal vulnerable code pattern at the GitHub scale, analyze its root causes, and show that it has contaminated popular LLMs. To the best of our knowledge, this code pattern was previously not studied in the literature. We received two CVEs for discovering this pattern in two very popular projects.

\section{Conclusions}\label{sec:conclusions}
We show that a critical path traversal vulnerability created by a single code pattern has polluted open-source projects on GitHub. We can detect, confirm, and remediate this vulnerability employing an automated pipeline that we developed. 
Using this pipeline at the GitHub scale, we have identified 1,756 vulnerable projects, and our responsible disclosure efforts led to 63 projects being patched. We have also shown that it might be difficult to fully eradicate this pattern globally, as popular LLMs seem to have been poisoned with it. In the future, we plan to improve and extend the pipeline in several directions, particularly by integrating other vulnerable code patterns and improving patch generation.


\begin{acks}
We thank Yury Zhauniarovich, our shepherd, and the reviewers for their helpful suggestions that allowed us to improve this paper. We also thank Tina Rezaei for optimizing the keyword selection algorithm for GitHub code search.

This research has been partially supported by the Dutch Research Council (NWO) under the project NWA.1215.18.008 Cyber Security by Integrated Design (C-SIDe).
\end{acks}

\bibliographystyle{ACM-Reference-Format}
\bibliography{references}


\appendix

\section{More Detailed Pipeline Results}\label{sec:more_details}
In this section, we provide additional analysis results from running the pipeline and specify our experimental setup. 

\begin{figure}[t]
  \centering
  \includegraphics[width=0.90\linewidth, keepaspectratio]{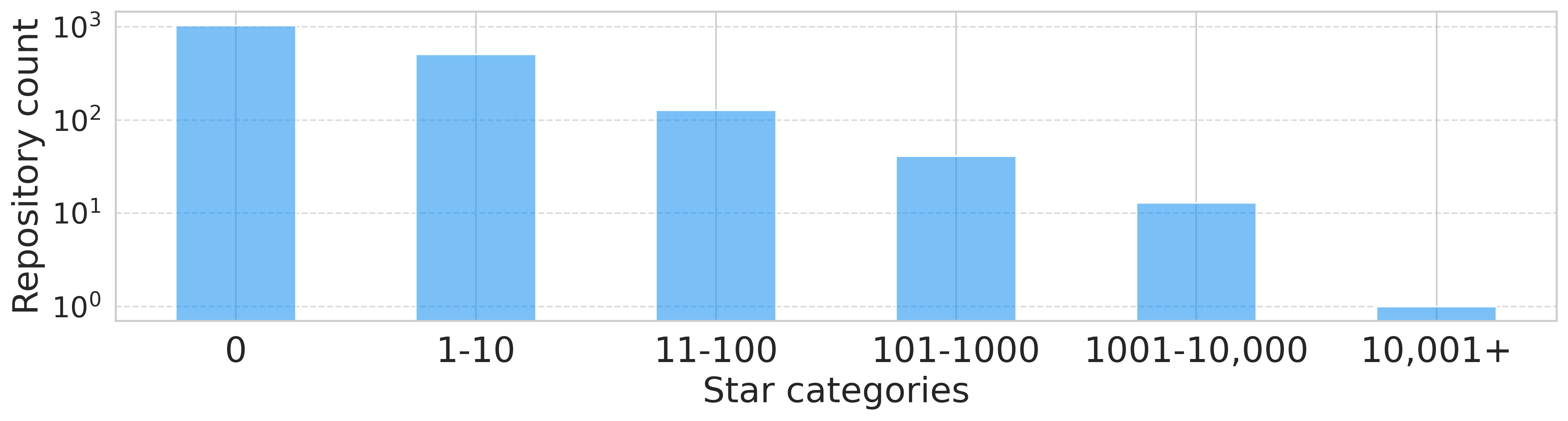}
  \caption{Popularity distribution of vulnerable projects (in the number of stars - log scaled).}
   \label{fig:starsdist}
\end{figure}

\textbf{Popularity of the affected projects.} GitHub stars are used as a metric to show popularity. Fig.~\ref{fig:starsdist} shows the distribution of vulnerable projects divided by the number of stars. It shows that the issue is not limited to less popular (or less seen) projects: the vulnerable code is spread into various projects, including some very popular ones. It can also be seen that the projects are not biased in a specific star range, as it is natural to have fewer projects with a high number of stars.

\begin{figure}[t]
  \centering
  \includegraphics[width=0.90\linewidth, keepaspectratio]{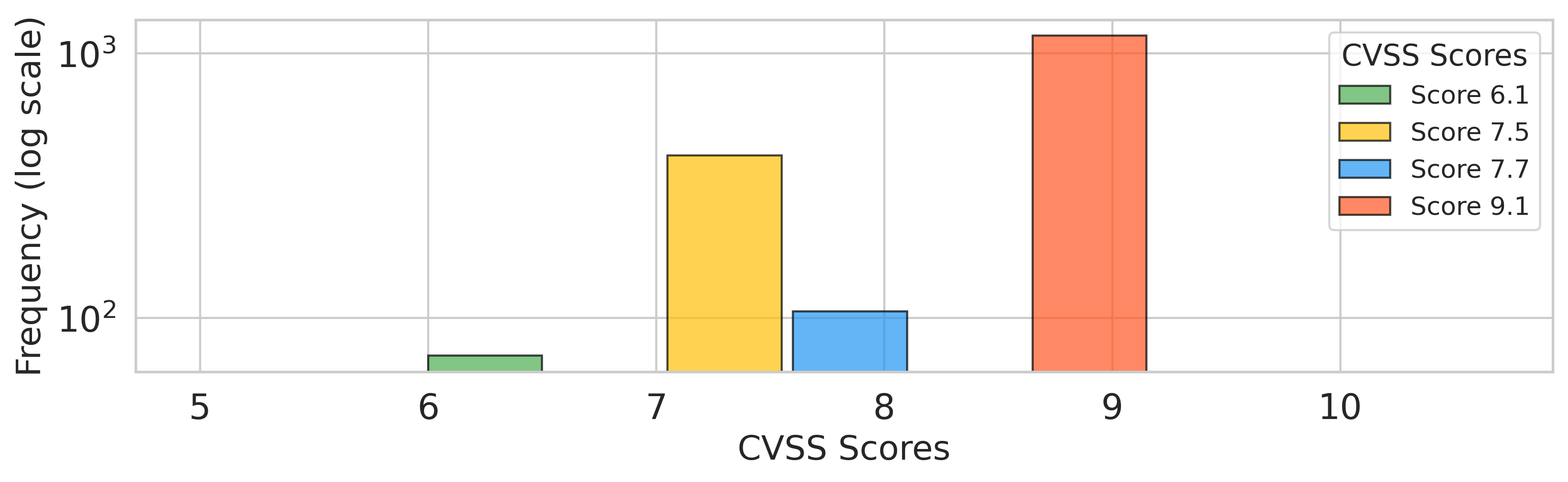}
  \caption{Automatically computed CVSS scores for exploited vulnerabilities.}
   \label{fig:cvssplot}
\end{figure}

\textbf{CVSS scores.} The distribution of CVSS scores for the confirmed vulnerabilities is shown in Fig.~\ref{fig:cvssplot}. As can be seen, the score ranges are between 6.1 and 9.1 (the median score is 9.1). This means that for most projects, the vulnerabilities are \emph{critical}: they are exploitable via network and vulnerable to denial-of-service attacks, thus the vulnerabilities pose a high risk if exploited.

\begin{figure}[t]
  \centering
  \includegraphics[width=0.90\linewidth, keepaspectratio]{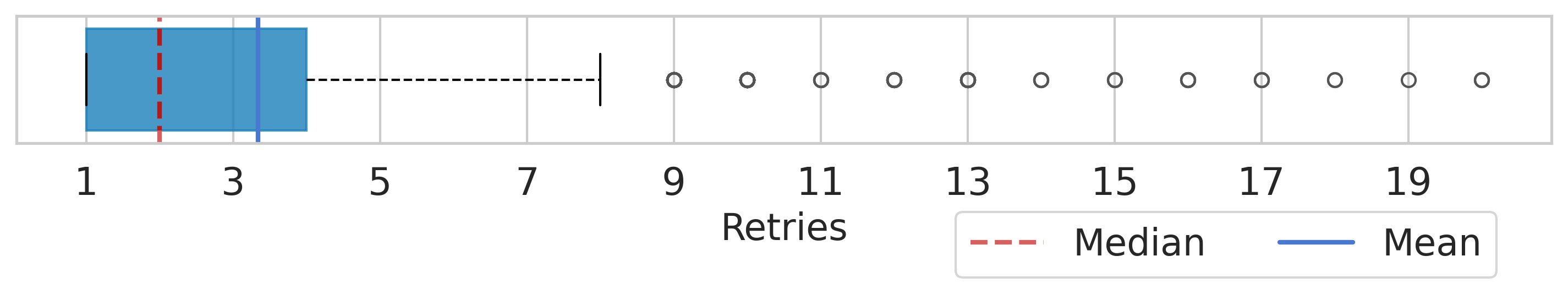}
  \caption{Number of LLM attempts for a successful patch of vulnerabilities.}
   \label{fig:llmtries}
\end{figure}

\textbf{Patch details.} The LLM does not always generate a correct patch on the first attempt (see Section~\ref{sec:llm-patcher}).
To quantify this, Fig.~\ref{fig:llmtries} reports the number of attempts required by the GPT-4 to patch a vulnerable project. As we can see, in most cases, the vulnerability was patched on the second try, but in some cases, it took more tries to fix it. 

Table~\ref{tab:patch_sizes} shows the summary statistics of the generated patch sizes. We can see that most projects received a patch with 5 added lines of code. We believe this is a relatively small patch that the maintainers can review and validate themselves. In 40 projects (out of 1600 with valid patches), the LLM has removed some lines of code. We inspected these patches separately. There, in 33 projects, the generated patch also added the removed lines (thus, they were simply moved); in 5 projects, the removed lines were added in a modified way (resulting in a more clean code); and in 2 projects these lines of code were removed: in one case, the patch removed commented out code (we still count this as removed LoCs); in the second case, the patch removed a logging code that would print the current date and the accessed file name in the developer console. This shows that the automated patching by the LLM mostly does not remove or modify the existing code functionality. Still, a review by the maintainers is necessary to fully understand the introduced code changes and assess them in the context of each project.

\begin{table}[t!]
  \centering
  \caption{Summary statistics of patch sizes. LoC$^+$ stands for lines of code added by the patch; LoC$^-$ stands for lines of code removed; LoC$^{t}$ is the total amount of modified lines of code (LoC$^+$+LoC$^-$). Note that each statistic is computed independently over all valid patches.} 
  \label{tab:patch_sizes}
  \begin{tabular}{lccc}
    \toprule
    \textbf{Statistic} & \textbf{LoC$^+$} & \textbf{LoC$^-$}  & \textbf{LoC$^t$}  \\
    \midrule
    Min & 1 & 0 & 1 \\
    Q1 & 5 & 0 & 5 \\
    Median & 5 & 0 & 5\\
    Q3 & 5 & 0 & 5 \\
    Max & 20 & 2 & 20 \\
     \bottomrule
  \end{tabular}
\end{table}   

\textbf{Our experimental setup.} 
We conclude with some statistics regarding the experimental setup to give an impression of the computing resources required for this study.
The whole experiment was conducted on a single server machine with a single AMD EPYC 7282 CPU @2.8 GHz with 64 GB of RAM running Ubuntu 22.04, and took approximately three weeks. The main bottleneck of the application was the GitHub Scraper due to API rate limits (see Section~\ref{sec:search}). The static analysis step took $\approx$3  seconds on average for each sample. In the Runner step, the main factor for the runtime duration is project size, as projects are cloned with Git, this step takes $\approx$15 seconds on average per project. The patching, which involves the use of GPT-4, takes $\approx$40 seconds on average, per project. Automatic validation and sending a pull request for a patch takes $\approx$54 seconds (however, this is not a bottleneck for our work, as we stage notifications). Within this procedure, first, several checks are performed: the current vulnerability status of the project is checked again, then if there was previously an issue opened on the issue, it is linked to the pull request, and also it double checks if there was a pull request sent previously to prevent duplicate pull requests. If everything checks out, the patch is committed, and the pull request is sent.

While the performance of the pipeline is practically feasible, the Runner step can be accelerated by locally caching projects and downloading them as a ZIP-file instead of cloning them with commit histories. As projects are isolated in containers, the steps are easily parallelizable, and this can be further extended by distributing projects over multiple nodes. Processing performance is not the main focus of this paper, we leave a discussion of performance improvements and scaling the pipeline to even larger scales for future work.

\begin{table}[t!]
  \centering
  \footnotesize
  \caption{Summary of successful exploit payloads}
  \label{tab:pocsstats}
  \begin{tabular}{>{\raggedright\arraybackslash}p{6cm}r}
    \toprule
    \textbf{Payload} & \textbf{Count} \\
    \midrule
    \texttt{../../../../../../flag.txt} & 1,749 \\
    \midrule
    \texttt{\%2e\%2e\%2f\%2e\%2e\%2f\%2e\%2e\%2f\%2fflag.txt} & 5 \\
    \midrule
    \texttt{images/../../../../../../../flag.txt} & 2 \\
    \bottomrule
  \end{tabular}
  \vspace{1mm}
  
  \par\smallskip
  \raggedright
  \footnotesize\textit{Note: Repeated patterns truncated for readability.}
\end{table}

\section{Survey Distributed to Maintainers}
\label{app:survey}
\noindent
\textbf{How would you rate the quality of the report? (Score from 1 to 10, where 10 is the highest)}
\begin{itemize}
    \item Overall report quality (Score: 1 to 10)
    \item Proof-of-Concept (PoC) code (Score: 1 to 10)
    \item CVSS metrics (Score: 1 to 10)
    \item Patch file (Score: 1 to 10)
    \item Vulnerability description (Score: 1 to 10)
\end{itemize}

\noindent
\textbf{Do you believe similar reports can be helpful for improving security?}

\begin{itemize}
    \item Strongly agree
    \item Somewhat agree
    \item Neither agree nor disagree
    \item Somewhat disagree
    \item Strongly disagree
\end{itemize}

\noindent
\textbf{What was your response to the vulnerability report?}

\begin{itemize}
    \item We believe it was vulnerable, and we have already applied the supplied patch.
    \item We believe it was vulnerable but we applied our patch.
    \item We believe it was vulnerable, but we didn’t fix it.
    \item The code was not in an important path, but we can still accept (or have already accepted) the patch as an improvement.
    \item It wasn’t vulnerable at all.
    \item I don’t know.
    \item No response.
\end{itemize}

\subsection{Maintainers’ responses to the survey}
\begin{figure}[h]
  \centering 
  \includegraphics[width=1.0\linewidth, keepaspectratio]{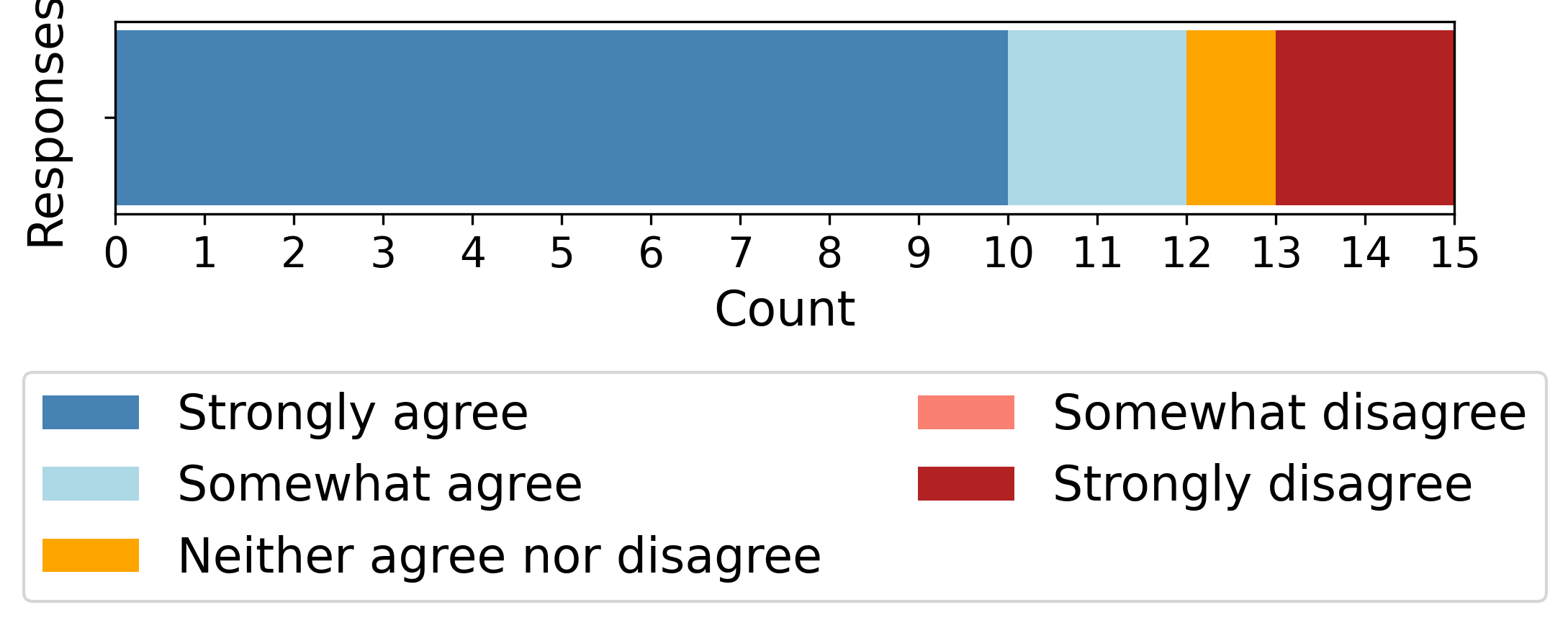}
  \caption{Responses from maintainers about the usefulness of similar security reports.}
   \label{fig:survey-usefulness}
\end{figure}

\begin{figure}[h]
  \centering
  \includegraphics[width=1.0\linewidth, keepaspectratio]{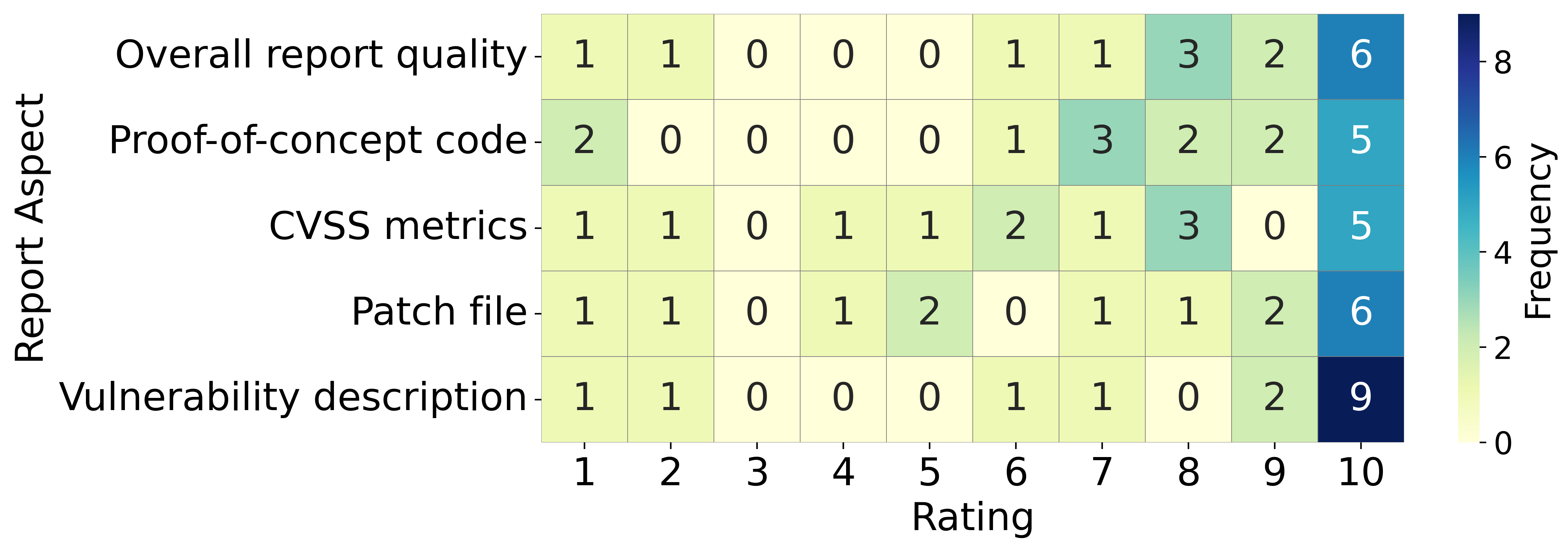}
  \caption{Heatmap of ratings for different parts of the report.}
   \label{fig:survey-rating}
\end{figure}

\begin{figure}[h]
  \centering
  \includegraphics[width=1.0\linewidth, keepaspectratio]{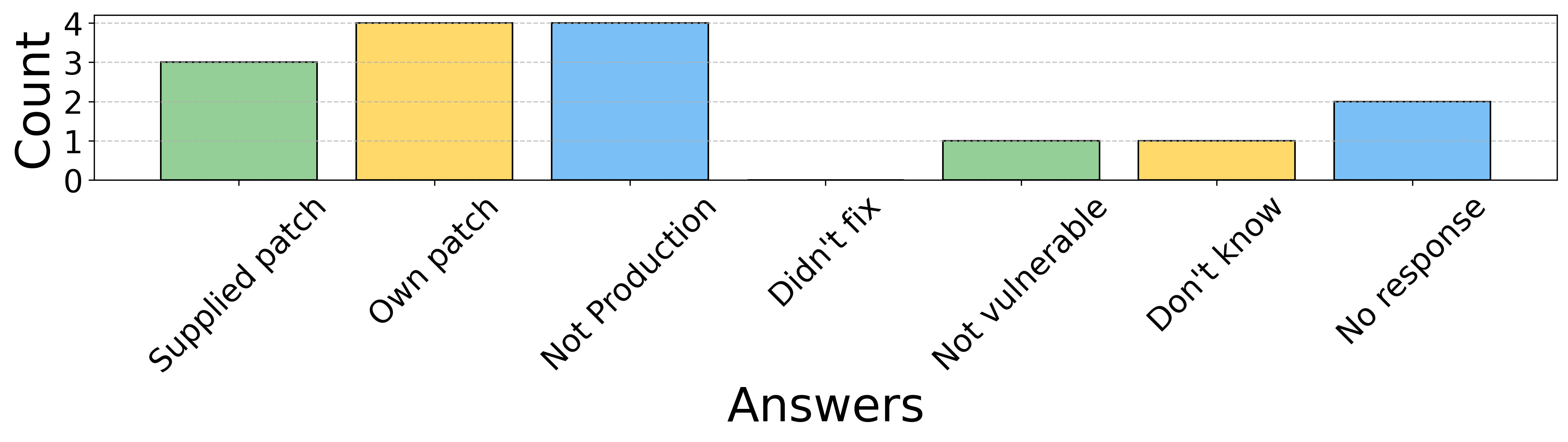}
  \caption{Developers' assessment of the vulnerability.}
   \label{fig:survey-action}
\end{figure}

\end{document}